\documentclass[12pt]{article}
\usepackage{graphicx}
\usepackage{amsmath}
\usepackage{multirow}
\usepackage{lineno}
\usepackage{booktabs}
\usepackage{placeins}

\newcommand\pubnumber{ATL-PHYS-PROC-2018-}

\newcommand\pubdate{\today}

\def\institute{Physikalisches Institut\\
Universit\"at Bonn, 53115 Bonn, Germany}

\def\support{\footnote{Supported by European Research Council grant
ERC-CoG-617185.}}

\makeatletter
\def\blfootnote{\xdef\@thefnmark{}\@footnotetext}
\makeatother

\def\copyright{\blfootnote{Copyright 2018 CERN for the benefit of the ATLAS Collaboration. CC-BY-4.0 license.}}

\def\Title#1{\begin{center} {\Large #1 } \end{center}}
\def\Author#1{\begin{center}{ \sc #1} \end{center}}
\def\Address#1{\begin{center}{ \it #1} \end{center}}

\newcommand\pubblock{\rightline{\begin{tabular}{l} \pubnumber\\
         \pubdate \end{tabular}}}
\newenvironment{Abstract}{\begin{quotation}  }{\end{quotation}}
\newenvironment{Presented}{\begin{quotation} \begin{center} 
             PRESENTED AT\end{center}\bigskip 
      \begin{center}\begin{large}}{\end{large}\end{center} \end{quotation}}
\def\Acknowledgements{\bigskip  \bigskip \begin{center} \begin{large}
             \bf ACKNOWLEDGEMENTS \end{large}\end{center}}

\def\beq{\begin{equation}}
\def\eeq#1{\label{#1}\end{equation}}
\def\eeqn{\end{equation}}

\def\beqa{\begin{eqnarray}}
\def\eeqa#1{\label{#1}\end{eqnarray}}
\def\eeqan{\end{eqnarray}}

\let\bar=\overbar

\def\Dslash{\not{\hbox{\kern-4pt $D$}}}
\def\dslash{\not{\hbox{\kern-2pt $\del$}}}

\def\alphas{\alpha_s}
\def\msb{{\bar{\ssstyle M \kern -1pt S}}}

\def\ttZ{\ensuremath{t \bar{t} Z}}

\def\ttW{\ensuremath{t \bar{t} W}}
\def\GeV{\ensuremath{\text{Ge\kern -0.1em V}}}
\def\TeV{\ensuremath{\text{Te\kern -0.1em V}}}

\begin{document}
\begin{titlepage}
\pubblock

\vfill
\Title{Top-quark pair production in association with a $W$ or $Z$ boson with the ATLAS experiment}
\vfill
	\Author{Sebastian Heer\support, on behalf of the ATLAS Collaboration\copyright}
\Address{\institute}
\vfill
\begin{Abstract}
	The cross section of the \ttZ~and \ttW~processes are measured in a
	simultaneous fit using 36.1 $\text{fb}^{-1}$ of proton-proton collisions at a centre of mass energy of $\sqrt{s}=13\ \TeV$ recorded by the ATLAS
	detector at the LHC. The result is found to be $\sigma_{t\bar{t}Z} = 0.95 \pm
	0.08 \text{ (stat.)} \pm 0.10 \text{ (syst.)}\,\text{pb}$ and $\sigma_{t\bar{t}W} =
	0.87 \pm 0.13 \text{ (stat.)} \pm 0.14 \text{ (syst.)}\,\text{pb}$ and
	compatible with the Standard Model.

\end{Abstract}
\vfill
\begin{Presented}
$11^\mathrm{th}$ International Workshop on Top Quark Physics\\
Bad Neuenahr, Germany, September 16--21, 2018
\end{Presented}
\vfill
\end{titlepage}
\def\thefootnote{\fnsymbol{footnote}}
\setcounter{footnote}{0}

\section{Introduction}

The \ttZ~and \ttW~processes provide direct access to the weak couplings of the
top quark. In addition, the two processes are often found to be a relevant
background in searches involving final states with multiple leptons and
$b$-quarks. Measurements of $\sigma_{\ttZ}$ and $\sigma_{\ttW}$ can be used to set
constraints on the weak couplings of the quarks, as well as impose limits on
possible beyond the Standard Model (BSM) effects. Previous measurements by the
ATLAS and CMS experiments~\cite{atlas,Chatrchyan:2008aa} indicate agreement with the Standard
Model~\cite{Aaboud:2016xve,Sirunyan:2017uzs}.

The dataset of 36.1 $\text{fb}^{-1}$ of proton-proton collisions at a centre of
mass energy of $\sqrt{s}=13\ \TeV$, collected in 2015 and 2016 by the ATLAS
detector is analysed. Final states with two, three or four isolated prompt
leptons are considered in the analysis. More details about the analysis can be
found in Reference~\cite{confnote}.

\section{Analysis Overview}

Separate analysis channels are defined to target the \ttZ~and \ttW~processes.
Each channel is divided into multiple regions in order to enhance the
sensitivity to the signal. Table~\ref{tab:intro-channels} lists the analysis
channels and the targeted decay modes of the \ttZ~and \ttW~processes.  

\begin{table}[htbp]
\centering
\caption{\label{tab:intro-channels} List of \ttW~and \ttZ~decay modes and
analysis channels targeting them. The symbols $b$ and $\nu$ denote a $b$-quark or antiquark and neutrino or antineutrino, respectively~\cite{confnote}.\vspace{1ex}}
\begin{tabular}{cccc}
\toprule
  Process & $t\bar{t}$ decay & Boson decay & Channel\\
\midrule
\multirow{2}{*}{$\ttW$}
&  $(\ell^{\pm}\nu b) (q\bar{q} b) $ & $\ell^{\pm}\nu$ & 2$\ell$ SS\\
& $ (\ell^{\pm}\nu b) (\ell^{\mp}\nu b)$ & $\ell^{\pm}\nu$ & Trilepton\\
\midrule
\multirow{2}{*}{\ttZ}
&  $(q\bar{q} b) (q\bar{q} b)$ & $\ell^{+}\ell^{-}$ & $2\ell$ OS\\
& $(\ell^{\pm}\nu b) (q\bar{q} b)$ & $ \ell^{+}\ell^{-}$ & Trilepton\\
& $(\ell^{\pm}\nu b) (\ell^{\mp} \nu b)$ & $ \ell^{+}\ell^{-}$ & Tetralepton\\
\bottomrule
\end{tabular}
\end{table}

The 2$\ell$-OS channel targets the \ttZ~process and is affected by large
backgrounds from $Z$+jets or $t\bar{t}$ production.  In order to separate the
signal from background in the most efficient way, Boosted Decision Trees (BDTs)
are used. Three BDTs are constructed and trained in regions requiring at least
5 jets. The number of input variables varies between 14 and 17. The most
important input variables are found to be the pseudorapidity of the dilepton
system and the transverse momenta of all jets divided by the sum of their
energies. Each signal region is divided into 19 equal sized bins of the BDT
discriminator. The normalization factors of the $Z$+jets background with one or
two heavy flavour jets, labeled as Z+1HF and Z+2HF, are determined in the fit
to data.

The 2$\ell$-SS channel targets the \ttW~process and uses a total of 12 signal
and control regions, classified by information about the lepton flavour, total charge and $b$-jet
multiplicity. The dominant background arises from fake leptons, which are
estimated by the matrix method~\cite{Aad:2010ey} and charge-flip leptons. The
control regions are used in a fit to measure fake and real lepton efficiencies,
as well as in the final fit to determine the signal cross sections,
anticorrelating the fake lepton background with the signal process. 

The regions in the 3$\ell$ channel are divided into two groups, depending on
whether a pair of OSSF leptons whose invariant mass is within $10\,\GeV$ of the
$Z$ boson mass is present. The signal regions are further categorized based on
jet and $b$-jet multiplicities. Four signal regions for each signal process are
defined. The dominant backgrounds for the regions targeting \ttZ~arise from
diboson production and the production of single top quarks in association with
a $Z$ boson. A control region is used to determine the normalization of the
$WZ$ background in the fit to data. The dominant background for the regions
targeting \ttW~arises from fake leptons, which is estimated by the matrix
method. 

The 4$\ell$ channel targets the \ttZ~process and uses four signal regions
depending on the $b$-jet multiplicity and the flavour of the leptons
from the $t\bar{t}$ system.  In addition, a control region is used to determine
the normalization of the $ZZ$ background in the fit to data. The dominant
backgrounds arise from the diboson production, the production of a single top
quark in association with a $W$ and a $Z$ boson, as well as fake leptons. The
fake lepton background is estimated from simulation, applying scale factors
obtained from a fit in control regions enriched by fake leptons.

\section{Results}

\begin{figure}[htbp]
\centering
\includegraphics[width=0.7\textwidth]{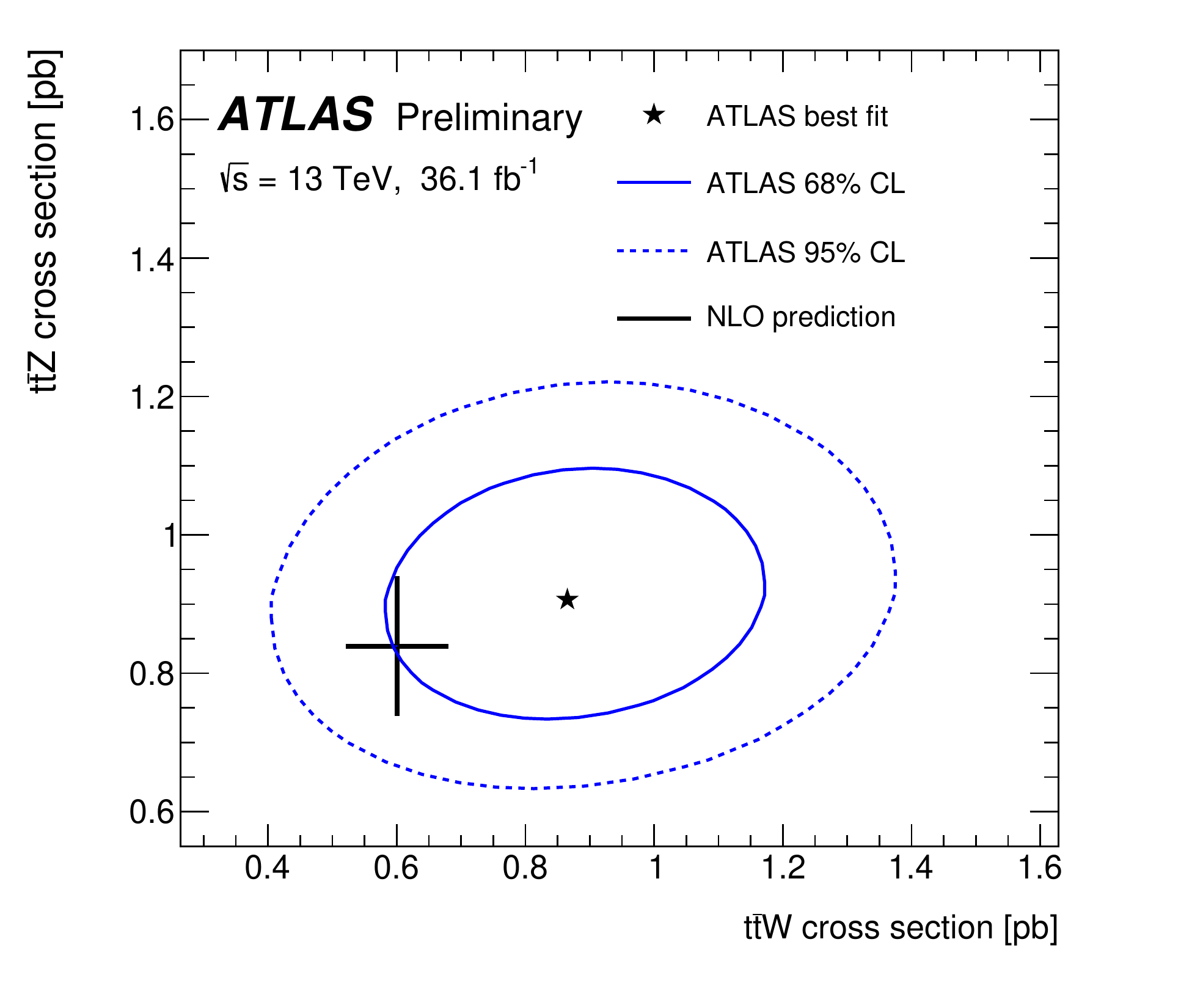}
\caption{\label{fig:2D_cross} The result of the simultaneous fit to the \ttZ~and \ttW~cross sections along with the 68\% and 95\% confidence level (CL)
	contours~\cite{confnote}. The cross shows the theoretical uncertainties in the SM
	predictions, and include renormalization and factorization scale
	uncertainties, as well as PDF uncertainties including $\alphas$ variations~\cite{Alwall:2014hca}.}
\end{figure}

The \ttZ~and \ttW~cross sections are simultaneously extracted using a binned
maximum-likelihood fit to the numbers of events in the dilepton, trilepton and
tetralepton signal and control regions. In total, the fit includes 11 signal
plus 2 control regions from the \ttZ, and 16 signal plus 12 control regions from
the \ttW~analysis. In all regions, good agreement between observed values and
the expectation is observed. The normalization corrections for the $WZ$ and
$ZZ$ backgrounds with respect to the predictions are obtained from the fit and
found to be compatible with unity: $0.91 \pm 0.10$ for the $WZ$ background and
$1.11 \pm 0.17$ for the $ZZ$ background.  The normalizations of the $Z$+1HF and
$Z$+2HF backgrounds are mainly constrained in the low BDT output bins of the
opposite-sign dilepton channel signal regions, where the signal contamination
is low. Their values are found to be $1.19\pm 0.25$ and $1.09\pm 0.13$,
respectively. The measured values of the cross sections from the combined fit
are $\sigma_{\ttZ} = 0.95 \pm 0.08 \text{ (stat.)} \pm
0.10 \text{ (syst.)}\,\text{pb} = 0.95 \pm 0.13 \,\text{pb}$ and $\sigma_{\ttW} =
0.87 \pm 0.13 \text{ (stat.)} \pm 0.14 \text{ (syst.)}\,\text{pb} = 0.87 \pm 0.19
\,\text{pb}$, assuming the SM value for the other cross section, respectively.
Figure~\ref{fig:2D_cross} shows a comparison of the fit results with
theoretical expectations, $\sigma_{\ttZ} = 0.88^{+0.09}_{-0.11}$ pb and
$\sigma_{\ttW} = 0.60^{+0.08}_{-0.07}$ pb~\cite{Alwall:2014hca}, demonstrating
good agreement between the measured and predicted cross sections. For the
\ttZ~process, both the observed and the expected significances are found to be
much larger than 5 standard deviations.  For the \ttW~process, an excess of
events over the expected Standard Model background-only hypothesis is found
with an observed (expected) significance of $4.3$ ($3.4$) standard deviations.
Table~\ref{tab:syst} shows the uncertainties in the measured \ttZ~and
\ttW~cross sections, grouped in categories, along with the total uncertainties.
For both processes, the precision of the measurement is affected by statistical
and systematic uncertainties in similar proportions.  For the
\ttZ~determination, the dominant systematic uncertainty sources are the
modelling of the backgrounds and of the signal.  For the \ttW~determination, the
dominant systematic uncertainty sources are the modelling of the signal and the
limited amount of data available in the control regions and simulated samples.

\begin{table}[htbp]
\centering \renewcommand{\arraystretch}{1.2}
\caption{List of relative uncertainties in the measured cross
sections of the \ttZ~and \ttW~processes from the fit, grouped in categories.
All uncertainties are symmetrized. The quadratic sum may not be equal to the total due to correlations between uncertainties introduced by the fit~\cite{confnote}.
\vspace{1ex}}
\label{tab:syst}
\begin{tabular}{lcc}
\toprule
Uncertainty                 &   $\sigma_{\ttZ}$ & $\sigma_{\ttW}$ \\
\midrule
Luminosity                                 &  2.9$\%$&   4.5$\%$\\
Simulated sample statistics         &  2.0$\%$&   5.3$\%$\\
Data-driven background statistics         &  2.5$\%$&   6.3$\%$\\
JES/JER                                    &  1.9$\%$&   4.1$\%$\\
Flavor tagging                            &  4.2$\%$&   3.7$\%$\\
Other object-related                       &  3.7$\%$&   2.5$\%$\\
Data-driven background normalization       &  3.2$\%$&   3.9$\%$\\
Modeling of backgrounds from simulation   &  5.3$\%$&   2.6$\%$\\
Background cross sections                  &  2.3$\%$&   4.9$\%$\\
Fake leptons and charge misID              &  1.8$\%$&   5.7$\%$\\
\ttZ modelling                             &  4.9$\%$&   0.7$\%$\\
\ttW modelling                             &  0.3$\%$&   8.5$\%$\\
\midrule
Total systematic            & 10$\%$ & 16$\%$\\
Statistical                 & 8.4$\%$& 15$\%$\\
\midrule
Total                       &  13$\%$& 22$\%$ \\
\bottomrule
\end{tabular}
\end{table}

\FloatBarrier

\Acknowledgements
The work of the author is currently funded by the European Research Council
under the European Union’s Seventh Framework Programme ERC Grant Agreement n.
617185.

\end{document}